\documentclass{article}
\usepackage{spconf,amsmath,graphicx}
\usepackage{fancyhdr}
\usepackage{hyperref}
\title{Comparison of SVD and Factorized TDNN approaches for speech to text}
\name{Jeffrey Josanne Michael \textsuperscript{*}, Nagendra Kumar Goel \textsuperscript{**}, Navneeth K \textsuperscript{**}, Jonas Robertson \textsuperscript{*}, Shravan Mishra \textsuperscript{*} \thanks{}}
\address{ \textsuperscript{*} Dialpad Inc.,  \textsuperscript{**} Go-Vivace Inc.}
\begin{document}
%\ninept
%
\maketitle
\begin{abstract}
 This work concentrates on reducing the RTF and word error rate of a hybrid HMM-DNN. Our baseline system uses an architecture with TDNN and LSTM layers. We find this architecture particularly useful for lightly reverberated environments. However, these models tend to demand more computation than is desirable. In this work, we explore alternate architectures employing singular value decomposition (SVD) is applied to the TDNN layers to reduce the RTF, as well as to the affine transforms of every LSTM cell. We compare this approach with specifying bottleneck layers similar to those introduced by SVD before training. Additionally, we reduced the search space of the decoding graph to make it a better fit to operate in real-time applications. We report -61.57\% relative reduction in RTF and almost 1\% relative decrease in WER for our architecture trained on Fisher data along with reverberated versions of this dataset in order to match one of our target test distributions.
\end{abstract}
\section{Introduction}
\label{sec:intro}
Sequence discriminative models such as those trained with the lattice free--maximum mutual information (LF-MMI) criterion are continuously improving with better deep learning strategies and more data from different distributions either by pooling strategies \cite{Yu:2010:ALS:1755252.1755412} or fine tuning \cite{8268947}. The \href{https://github.com/kaldi-asr/kaldi/blob/eacf34a85ab7ece6a76bd73b9443bc2fe62ac6f1/egs/aspire/s5/local/chain/tuning/run_tdnn_lstm_1a.sh#L218}{state-of-the-art} hybrid architecture comprises bi-directional long short term Memory (BLSTM) neural networks used together with time-delay neural networks (TDNNs) stacked one over the other. This hybrid ASR setup still requires complementary help from Hidden Markov Models (HMMs), which are responsible for aligning the sequence triphones with the variable length audio data. However, these models are difficult to use in a real-time setting due to the very large number of parameters involved in LSTMs. To overcome this barrier to using an LSTM-TDNN architecture in real-time, we compare and contrast parameter reduction approaches such as SVD and its comparably equivalent bottleneck architectures.
% \fancypagestyle{alim}
% {\fancyhf{}\renewcommand{\headrulewidth}{0pt}\fancyfoot[L]{\textsuperscript{1} \href{https://github.com/kaldi-asr/kaldi/blob/eacf34a85ab7ece6a76bd73b9443bc2fe62ac6f1/egs/aspire/s5/local/chain/tuning/run_tdnn_lstm_1a.sh#L218}{LSTM-TDNN hybrid} }}
% \thispagestyle{alim}

\section{ Background and Related Work}
\label{sec:format}
% Hybrid systems are never trained from scratch but they are bootstrapped by using time alignments from HMM-GMM systems \cite{Morgan1995AnIT}.
% The final model, for inference, is obtained after composing the acoustic model, statistical language model and a pronunciation model.
% State-of-the-art performance is obtained by injecting more prior information from a generative model (HMM-GMM), formerly the state-of-the-art approach \cite{6296526}, to the Neural nets. The prior information is the time alignments for every spoken phone in an utterance. For a DNN, this information is very vital as the ground truth will be compared against the hypothesis tokens every frame. 
Recent acoustic models have been trained using sequence discriminative cost functions without cross-entropy pre-training \cite{Povey2016PurelySN}. Different types of sequence-discriminative cost functions are MMI (Maximum Mutual Information), bMMI (boosted-MMI), MPE (Minumum Phone Error), MBR (Minimum Bayes Rate) and sMBR (state-level MBR) \cite{mmi_mbr}. In our work, we use MMI as the cost function for all of the architectures, though cross entropy is used for regularization.

% Cross-entropy pre-training is used to generate denominator lattices which are used for sequence discriminative training . HMM-DNN models trained without frame-level pretraining, use purely sequence discriminative criterion, for example LF-MMI \cite{Povey2016PurelySN}. DNNs are trained with a large number of output states. This in turn leads to large number of parameters, mostly concentrated in the final layer. Few output targets are actually active and out of those that are active, they are probably correlated. It is suspected that the last weight layer is of low rank, which means factorization can be used to reduce the number of parameters significantly \cite{6638949}.

% Our work differs from \cite{6638949} in the application of factorization to hidden layers, not just the final weight layer, and then stabilizing the network. Our approach is described in detail in Section \ref{sec:arch}.

Though the main focus of our work is latency reduction, one could consider it similar to other model compression techniques proposed in \cite{6638949} and \cite{Nakkiran2015CompressingDN}. \cite{6638949} and \cite{Nakkiran2015CompressingDN} use low-rank factorization for bottlenecking but this technique differs from our work in the way bottlenecking is applied. Also, techniques in \cite{6638949} are effective only when bottlenecks are deployed in the input and output layers.
And authors in \cite{Nakkiran2015CompressingDN} apply model compression techniques to a Keyword Search (KWS) task whereas in our work we concentrate on ASR.

\subsection{TDNN and its factorized variant}
\label{ssec:tdnnf}
TDNN architecture progressively captures more temporal context as the layers get deeper.
 The temporal modelling at every layer is controlled by two factors: subsampling and splicing \cite{Peddinti2015ATD}.
 In the TDNN-F (factorized) variant \cite{inproceedingstdnnf}, the layers are modified in two ways. In addition to factorization, highway layers are introduced, which connect non-adjacent layers. This contrasts with \cite{Nakkiran2015CompressingDN}, where layers are only factorized. Layer factorization is done using SVD. Factorization bottlenecks the affine transformation matrix (weight matrix). This reduced weight matrix has lower dimensionality. There are two side effects of factorization: while there is an improvement in computational efficiency, as the dimensionality is reduced, there is also a reduction in representational expressiveness. Unfortunately, instability is introduced by pruning away dimensions at every layer and by randomly initializing the weight matrices.
%  SVD helps in effective data representation by selecting features containing valuable temporal feature information. 

 To stabilize training, the highway layers from \cite{srivastava2015highway} are used. Highway layers are not standalone layers, but when applied to an existing layer, they forward the output dimensions to a deeper layer by skipping one or more immediately following layers. These layers are used in scenarios where information loss occurs at deeper layers. By skipping one or more layers, unimpeded information flows across several deeper layers. This also helps avoid vanishing gradients in backpropagation through very deep layers. Authors from \cite{inproceedingstdnnf} claim that the TDNN-F models perform about as well as LSTM-TDNN hybrids. However, from our experiments in Table \ref{main_table}, the claim did not hold; hence, we continued to explore other possibilities. 
\begin{table*}[!ht]
\label{tab1}

\begin{tabular}{|l|l|l|}
\hline
\textbf{Model} & \textbf{Splicing indices} & \textbf{Number of parameters} \\ \hline
Baseline & \begin{tabular}[c]{@{}l@{}}\{(-2,-1,0,1,2) , (-1,0,1) , (-1,0,1) , LSTM , (-3,0,3) , (-3,0,3) , LSTM , (-3,0,3) , (-3,0,3) \\  , LSTM , (-3,0,3) , (-3,0,3) , LSTM\}\end{tabular} & 49,945,168 \\ \hline
LSTM-TDNN-I & \begin{tabular}[c]{@{}l@{}}\{(-1,0,1) , (-1,0,1) , (-1,0,1) , LSTM , (-3,0,3) , (-3,0,3) , LSTM , (-3,0,3) , (-3,0,3)\\   , LSTM , (-3,0,3) , (-3,0,3) , LSTM\}\end{tabular} & 49,904,288 \\ \hline
LSTM-TDNN-II & \begin{tabular}[c]{@{}l@{}}\{(-2,-1,0,1,2) , (-1,0,1) , (-1,0,1) , LSTM , (-3,0,3) , (-3,0,3) , LSTM , (-3,0,3) , (-3,0,3) \\  , LSTM , (0) , (0) , LSTM\}\end{tabular} & 46,840,480 \\ \hline

\end{tabular}
\caption{Chosen splicing indices for different architectural variants}
\end{table*}

% \subsection{LSTM-TDNN-F}
% \label{ssec:lstmtdnnf}
% Just the TDNN-F architecture itself, without any recurrent layers, yielded poorer WER results compared to LSTM+TDNN
% architecture, which yielded poorer RTF compared to TDNN-F architecture.
% As explained in the previous section, the factorized TDNN layers in a TDNN-F architecture are computationally less intensive than the TDNN layers in a LSTM+TDNN
% architecture. Due to the fact that matrices are being propagated by skipping intermediate layers, the matrices become sparser with depth. This results in the sparsity of TDNN-F layers affecting deeper layers. This loss of information due to sparsity explains the poorer WER performance.

% The idea of an LSTM-TDNN-F architecture is appealing; however, the issue of TDNN-F layers passing information to interleaved LSTM layers, leading to more dimensions in recurrent cells, is a drawback as it impacts  latency which is important in a real-time system deployed to a production environment.
% To get the best of both worlds we decided to have TDNN layers bottlenecked using SVD, which resulted in an LSTM+TDNN(SVD) architecture with the dimensional sparsity issue ameliorated due to the LSTM layers.

\subsection{Frame splicing}
\label{ssec:lfr}
TDNN, as discussed in section \ref{ssec:tdnnf}, captures narrow windows at the initial layers and complete input at the deeper layers. From \cite{7968318}, we derive two architectures by tweaking the frames that are spliced at different layers which are described in  table \ref{tab1}. For our first architecture (LSTM-TDNN-I), we changed the frame splicing from \{-2,-1,0,+1,+2\} to \{-1,0,1\}. Our second choice (LSTM-TDNN-II) was to architect a model with a lower splice window (\{ -3, 0, +3\} to \{0\}) at the deeper layers. Since traditionally we do not splice frames to the LSTM layers, we deployed different splices only before the TDNN layers. Our first architecture (LSTM-TDNN-I) reduced the number of parameters from 49,945,168 to 49,904,288 (reduction of 40,880 parameters). The second choice resulted in 46,840,480 parameters, from 49,945,168 (reduction of 3,104,688).

\section{architecture: LSTM-TDNN-SVD}
\label{sec:arch}
\subsection{LSTM-TDNN-SVD}
\label{ssec:stable}

Training an LSTM-TDNN-SVD network is a two-part process. The standard LSTM-TDNN architecture is first trained from the work\textsuperscript{1} until its convergence. Then the model is compressed by applying SVD \cite{doi:10.1137/1035134} to all of the affine transformations in the layers, a process that is controlled by two parameters. To decompose a matrix, the weight matrix of the affine layer (before applying ReLU non-linearity) is forced to be an identity matrix when multiplied with its transpose. The singular values and two Eigen matrices are produced as the result:\\
\begin{displaymath}  SVD: Wx =  \omega \epsilon \upsilon \end{displaymath} where $\displaystyle \omega,  \upsilon $ are Eigen matrices, and  $\displaystyle  \epsilon $ is the singularity matrix.
 
\begin{figure}[h]
\includegraphics[width=8cm]{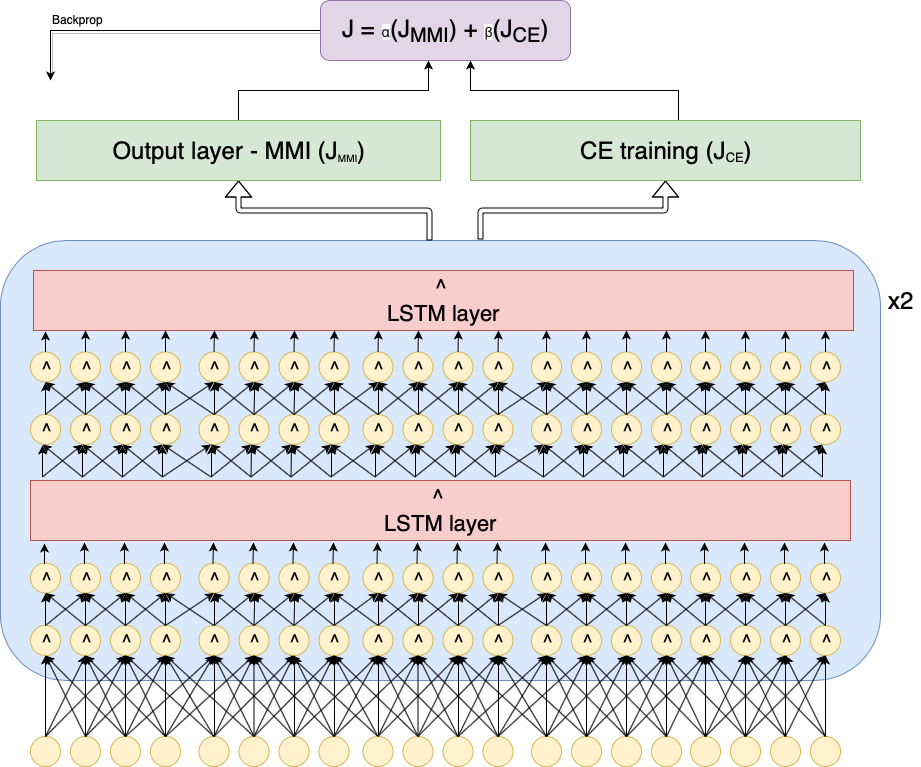}
\label{fig:tdnnsvd}
\caption{Figure. LSTM-TDNN-SVD architecture}

\end{figure}

For matrix E, only the singular values which contribute to energy threshold times the total energy of the singularity parameters are retained. These singularity parameters are sorted in descending order, and lower values are pruned out until the total energy (sum of squares) of the pruned set of parameters is just above (threshold * total energy). The values which are pruned away are replaced with 0 in the singularity matrix (E), and the weights matrix after SVD is derived with shrunken dimensions. Also, if the shrinkage ratio of the SVD refactored weights matrix is higher than the shrinkage threshold for any of the TDNN layers, the SVD process is omitted for that particular affine weights layer. This omission of SVD did not happen at any of the layers in our architecture. The architecture of stacked TDNN and LSTM layers, which we call "LSTM-TDNN-SVD", can be found in Figure \ref{fig:tdnnsvd}.  The \^{} in the figure represent the application of SVD at respective layers.
% \vfill
\subsection{Stabilized bottlenecked-LSTM-TDNN}
\label{ssec:stable}
Instead of adopting the approach of \cite{inproceedingstdnnf} to achieve stability in training, we achieve convergence in training by interleaving LSTM layers in the bottlenecked architecture. In our approach, we start the training from scratch by randomly initializing the weight matrices. The configuration of this architecture along with the number of output dimensions are the same as those from the ”LSTM-TDNN-SVD” architecture. We would like to emphasize the fact that no SVD is happening in this approach; only SVD dimensions are being copied. LSTM layers stabilize the training because the LSTM cells retain the important dimensions that carry valid temporal information before feeding their matrices to the next bottlenecking TDNN layer.

% \begin{minipage}[b]{1.0\linewidth}
%   \centering
%   \centerline{\includegraphics[width=8.5cm]{tdnnb-lstm-4.png}}
%   \label{fig:tdnnsvd}
%  \vspace{0.5cm}
% %   \centerline{ LSTM-TDNN-SVD architecture}\medskip
% \caption{Figure. LSTM-TDNN-SVD architecture}
% \end{minipage}

\begin{table*}[!ht]

\label{main_table}
\begin{center}
\begin{tabular}{|l|l|l|l|}
\hline
\textbf{Model}    & \textbf{parameters} & \textbf{RTF}\textsuperscript{2} & \textbf{WER} \\ \hline
TDNN              &      36,803,440            &        0.687     &    24.83           \\ \hline
LSTM + TDNN       &      49,945,168            &       1.132       &      17.89         \\ \hline
LSTM + TDNN - I    &        49,904,288          &       0.8962       &       21.1088        \\ \hline
LSTM + TDNN - II   &      46,840,480            &       0.9762       &       24.623        \\ \hline
% LSTM + TDNN-F      &    [\ref{ssec:lstmtdnnf}]               & [\ref{ssec:lstmtdnnf}]       &    [\ref{ssec:lstmtdnnf}]            \\ \hline
TDNN-F             &      6,218,656            &        0.28      &       22.235        \\ \hline
TDNN - SVD        &           \cite{inproceedingskws}      &     ~ \cite{inproceedingskws}        &      (training instability)         \\ \hline
LSTM - TDNN - SVD + active\_states=5000 &        22,716,596          &         0.67     &       18.100        \\ \hline
LSTM - TDNN - SVD + active\_states=500 &    22,716,596              &         0.435     &     18.324          \\ \hline
LSTM - TDNN - SVD + random init. &    22,716,596              &         0.68     &     17.723          \\ \hline
\end{tabular}
\end{center}
\caption{Different architectures and their corresponding metrics}
\end{table*}
\begin{table*}[!ht]

\label{in_n_out}
\begin{tabular}{|l|l|l|l|l|l|l|l|l|l|l|l|l|l|l|}
\hline
                    & \textbf{TD1} & \textbf{TD2} & \textbf{TD3} & \textbf{LS1} & \textbf{TD4} & \textbf{TD5} & \textbf{LS2} & \textbf{TD6} & \textbf{TD7} & \textbf{LS3} & \textbf{TD8} & \textbf{TD9} & \textbf{LS4} & \textbf{Out} \\ \hline
\textbf{Input dim}  & 300          & 3072         & 3072         & 1280         & 1536         & 3072         & 1280         & 1536         & 3072         & 1280         & 1536         & 3072         & 1280         & 512          \\ \hline
\textbf{Output dim} & 1024         & 1024         & 1024         & 512          & 1024         & 1024         & 512          & 1024         & 1024         & 512          & 1024         & 1024         & 512          & 5968         \\ \hline
\textbf{SVD'd Out}  & 127          & 319          & 317          & 406          & 183          & 372          & 404          & 164          & 385          & 427          & 173          & 433          & 501          & 5968         \\ \hline
\end{tabular}
\caption{Study of dimensions output by SVD}
\end{table*}

\section{Experimental Setup}
\label{sec:majhead}
Experiments were conducted using the Kaldi toolkit \cite{Povey_ASRU2011}. We report a set of results for various models trained on the Fisher data set \cite{article_fischer} along with its reverberated versions.

\subsection{Data}
Fisher \cite{article_fischer} and its reverberated versions were used for training. The audio data were reverberated as one of our target test distributions typically occur in reverberated ambience. The development set and test set were curated by collecting audio data of in-domain conversational telephone speech (CTS). The work was concentrated towards using this system for this domain, hence the choice of novel test and development sets.
% From, various work \cite{unknown_mmi}, Maximum mutual information or its variant is used as the training criterion and trained in a sequence-discriminative fashion. Our acoustic model is comprised of 5968 senones. . The Fisher data set was used for training along with its four reverberated counterparts. Several regularization techniques were used, including a dropout schedule, which dropped several input and hidden nodes. The dropout schedule used was {0.0, 0.2, 0.3, 0.5, 0.0}.

% \subsection{Hardware details}
% A Google Compute Engine instance was used for training the models. This virtual machine was allocated Four Tesla P100 GPUs with 120 MB of RAM, which accounts for the 64 mini-batches that were used in this training.

\fancypagestyle{alim}{\fancyhf{}\renewcommand{\headrulewidth}{0pt}\fancyfoot[L]{\textsuperscript{2} RTF= (total time taken by the system) / (total duration of the audio)}}
\thispagestyle{alim}

\subsection{Experiment details}
This work focused on obtaining the best WER possible while at the same time maintaining the real-time constraints of our system in production. Thus, the LSTM-TDNN architecture with the highest RTF was considered as the baseline. 'LSTM+TDNN-I' from Table \ref{main_table} was architected by modelling a narrower window at the initial layers and maintaining the frame rate at the deeper layers. The other model had a single frame input at the deeper layers by having a wider context window at the initial layers. Though these architectures were chosen primarily to reduce the real-time factor, previous work \cite{7968318} has shown that the low-frame-rate architectures outperformed the considered baseline models in both RTF and WER. However, the reduced RTF was still not ideal for our real-time environment. The proposed architecture (LSTM-TDNN-SVD) was also evaluated by training on the same data, and we report our observations in Table \ref{main_table}. The architecture reduced the real-time factor by 40.99\% relative to the baseline model. We performed more decoder parameter tuning to further reduce the RTF to 61.57\% relative as reported in Table \ref{main_table}.

\subsection{Observation and Results}
\label{ssec:obs}
In this section we share our intuition about the observations seen in various experiments performed by us.

In Table 2 for the LSTM-TDNN-SVD models, reducing the active states from 5000 to 500 gives a reasonable reduction in RTF with a tolerable increase in WER. `LSTM-TDNN-SVD+random init' yields optimum WER and RTF.

We also found one other interesting observation. Though SVD is used here to reduce the latency of the larger model, SVD implicitly retains the important dimensions, and we prune out the dimensions of lesser temporal information. In Table \ref{in_n_out}, for the layers TD2, TD3, TD5, TD7, TD9 which have the same input dimension '3072', the reduced SVD dimensions generally increased: 319$\rightarrow$ 317$\rightarrow$ 372$\rightarrow$ 385$\rightarrow$ 433. This trend confirms that the instability in a purely factorized but non-highway TDNN layer training is likely stabilized by the LSTM layers, as the increase in important features take place only after the interleaving LSTM layers and not between two LSTM layers.

% \section{DATADOG ILLUSTRATIONS}
% \label{sec:illust}

\section{Conclusion}
\label{sec:conclusion}
In conclusion, SVD during training gives us the optimum number of output dimensions, although that does not necessarily mean that WER will be optimal. However, the SVD-obtained output dimensions can further be explored in future experiments. For example, we could round them to nearest multiple of eight or four and train the models from scratch. This technique is not only expected to reduce training time but based on our current results and observations it would also likely improve WER.
Now that we know that SVD-inspired linear bottleneck layers can train from scratch, we have started multiple training experiments for different variants of output dimensions. We will include additional results after this paper's acceptance and submit as the final paper.

\section{Acknowledgement }
\label{sec:acknowledgement}

% Below is an example of how to insert images. Delete the ``\vspace'' line,
% uncomment the preceding line ``\centerline...'' and replace ``imageX.ps''
% with a suitable PostScript file name.
% -------------------------------------------------------------------------
% \begin{figure}[htb]

% %
% \begin{minipage}[b]{.48\linewidth}
%   \centering
%   \centerline{\includegraphics[width=4.0cm]{image3}}
% %  \vspace{1.5cm}
%   \centerline{(b) Results 3}\medskip
% \end{minipage}
% \hfill
% \begin{minipage}[b]{0.48\linewidth}
%   \centering
%   \centerline{\includegraphics[width=4.0cm]{image4}}
% %  \vspace{1.5cm}
%   \centerline{(c) Result 4}\medskip
% \end{minipage}
% %
% \caption{Example of placing a figure with experimental results.}
% \label{fig:res}
% %
% \end{figure}

%\vfill
%\pagebreak

% \vfill\pagebreak

% \section{REFERENCES}
% \label{sec:refs}

\bibliographystyle{IEEEbib}
\bibliography{strings,refs}

\begin{thebibliography}{10}

\bibitem{Yu:2010:ALS:1755252.1755412}
Dong Yu, Balakrishnan Varadarajan, Li~Deng, and Alex Acero,
\newblock ``Active learning and semi-supervised learning for speech
  recognition: A unified framework using the global entropy reduction
  maximization criterion,''
\newblock {\em Comput. Speech Lang.}, vol. 24, no. 3, pp. 433--444, July 2010.

\bibitem{8268947}
P.~{Ghahremani}, V.~{Manohar}, H.~{Hadian}, D.~{Povey}, and S.~{Khudanpur},
\newblock ``Investigation of transfer learning for asr using lf-mmi trained
  neural networks,''
\newblock in {\em 2017 IEEE Automatic Speech Recognition and Understanding
  Workshop (ASRU)}, Dec 2017, pp. 279--286.

\bibitem{Povey2016PurelySN}
Daniel Povey, Vijayaditya Peddinti, Daniel Galvez, Pegah Ghahremani, Vimal
  Manohar, Xingyu Na, Yiming Wang, and Sanjeev Khudanpur,
\newblock ``Purely sequence-trained neural networks for asr based on
  lattice-free mmi,''
\newblock in {\em INTERSPEECH}, 2016.

\bibitem{mmi_mbr}
Karel Veselý, A.~Ghoshal, Lukas Burget, and Daniel Povey,
\newblock ``Sequence-discriminative training of deep neural networks,''
\newblock {\em Proceedings of the Annual Conference of the International Speech
  Communication Association, INTERSPEECH}, pp. 2345--2349, 01 2013.

\bibitem{6638949}
T.~N. {Sainath}, B.~{Kingsbury}, V.~{Sindhwani}, E.~{Arisoy}, and
  B.~{Ramabhadran},
\newblock ``Low-rank matrix factorization for deep neural network training with
  high-dimensional output targets,''
\newblock in {\em 2013 IEEE International Conference on Acoustics, Speech and
  Signal Processing}, May 2013, pp. 6655--6659.

\bibitem{Nakkiran2015CompressingDN}
Preetum Nakkiran, Raziel Alvarez, Rohit Prabhavalkar, and Carolina Parada,
\newblock ``Compressing deep neural networks using a rank-constrained
  topology,''
\newblock in {\em INTERSPEECH}, 2015.

\bibitem{Peddinti2015ATD}
Vijayaditya Peddinti, Daniel Povey, and Sanjeev Khudanpur,
\newblock ``A time delay neural network architecture for efficient modeling of
  long temporal contexts,''
\newblock in {\em INTERSPEECH}, 2015.

\bibitem{inproceedingstdnnf}
Daniel Povey, Gaofeng Cheng, Yiming Wang, Ke~Li, Hainan Xu, Mahsa Yarmohammadi,
  and Sanjeev Khudanpur,
\newblock ``Semi-orthogonal low-rank matrix factorization for deep neural
  networks,''
\newblock 09 2018, pp. 3743--3747.

\bibitem{srivastava2015highway}
Rupesh~Kumar Srivastava, Klaus Greff, and Jürgen Schmidhuber,
\newblock ``Highway networks,'' 2015.

\bibitem{7968318}
V.~{Peddinti}, Y.~{Wang}, D.~{Povey}, and S.~{Khudanpur},
\newblock ``Low latency acoustic modeling using temporal convolution and
  lstms,''
\newblock {\em IEEE Signal Processing Letters}, vol. 25, no. 3, pp. 373--377,
  March 2018.

\bibitem{doi:10.1137/1035134}
G.~W. Stewart,
\newblock ``On the early history of the singular value decomposition,''
\newblock {\em SIAM Review}, vol. 35, no. 4, pp. 551--566, 1993.

\bibitem{inproceedingskws}
George Tucker, Minhua Wu, Ming Sun, Sankaran Panchapagesan, Gengshen Fu, and
  Shiv Vitaladevuni,
\newblock ``Model compression applied to small-footprint keyword spotting,''
\newblock 09 2016, pp. 1878--1882.

\bibitem{Povey_ASRU2011}
Daniel Povey, Arnab Ghoshal, Gilles Boulianne, Lukas Burget, Ondrej Glembek,
  Nagendra Goel, Mirko Hannemann, Petr Motlicek, Yanmin Qian, Petr Schwarz, Jan
  Silovsky, Georg Stemmer, and Karel Vesely,
\newblock ``The kaldi speech recognition toolkit,''
\newblock in {\em IEEE 2011 Workshop on Automatic Speech Recognition and
  Understanding}. Dec. 2011, IEEE Signal Processing Society,
\newblock IEEE Catalog No.: CFP11SRW-USB.

\bibitem{article_fischer}
Christopher Cieri, David Miller, and Kevin Walker,
\newblock ``The fisher corpus: A resource for the next generations of
  speech-to-text,''
\newblock 01 2004.

\end{thebibliography}

\end{document}